\documentclass{article}

\usepackage{arxiv}
\usepackage[square,numbers,sort&compress]{natbib}

\usepackage[utf8]{inputenc} 
\usepackage[T1]{fontenc}    
\usepackage{hyperref}       
\usepackage{url}            
\usepackage{booktabs}       
\usepackage{amsfonts}       
\usepackage{nicefrac}       
\usepackage{microtype}      
\usepackage{lipsum}		
\usepackage{graphicx}

\usepackage{doi}
\usepackage{multirow}
\usepackage{amsmath}
\usepackage{pgfplots}
\usepackage{xcolor}
\usepackage{geometry}
\usepackage{graphicx} 
\usepackage{booktabs}
\pgfplotsset{compat=1.18}
\usepackage{listings}
\title{LLMs for Engineering: Teaching models to design High Powered Rockets}

\author{
Toby Simonds \\
Tufa Labs\\
\texttt{toby@tufalabs.ai} \\
}

\renewcommand{\shorttitle}{\textit{arXiv} Template}

\hypersetup{
pdftitle={A template for the arxiv style},
pdfsubject={q-bio.NC, q-bio.QM},
pdfauthor={Toby s},
pdfkeywords={First keyword, Second keyword, More},
}

\begin{document}
\maketitle
\begin{abstract}

Large Language Models (LLMs) have transformed software engineering, but their application to physical engineering domains remains underexplored. This paper evaluates LLMs' capabilities in high-powered rocketry design through RocketBench, a benchmark connecting LLMs to high-fidelity rocket simulations. We test models on two increasingly complex design tasks: target altitude optimization and precision landing challenges. Our findings reveal that while state-of-the-art LLMs demonstrate strong baseline engineering knowledge, they struggle to iterate on their designs when given simulation results and ultimately plateau below human performance levels. However, when enhanced with reinforcement learning (RL), we show that a 7B parameter model outperforms both SoTA foundation models and human experts. This research demonstrates that RL-trained LLMs can serve as effective tools for complex engineering optimization, potentially transforming engineering domains beyond software development.

\end{abstract}

\section{Introduction}
Large Language Models (LLMs) have significantly transformed software engineering practices, yielding quantifiable improvements in code generation, debugging processes, and documentation development. Studies demonstrate productivity enhancements of 55\% on various software development task completion \citet{GitHub}. Despite these documented efficiency gains in computational domains, LLMs' applications to mechanical, aerospace, civil, and other physical engineering disciplines remain underdeveloped. This disparity raises a fundamental research question: Can LLMs function as effective tools for engineering tasks beyond software development?

To explore this hypothesis, we selected high-powered rocketry as an ideal test domain for several compelling reasons. First, rocket design represents a well-bounded yet complex engineering challenge with clear performance metrics. Second, the physics of rocketry incorporates multiple engineering disciplines (aerodynamics, structural mechanics, propulsion systems) while remaining computationally tractable. Third, the domain offers a straightforward way to quantify success through metrics such as altitude accuracy and landing precision. Finally, model rocketry provides a practical analogue to more complex aerospace engineering challenges while remaining accessible for research purposes.

This paper presents an interface enabling LLMs to design model rockets using RocketPy\citep{RocketPy}, a high-fidelity trajectory simulation tool. We evaluate these models on rocket design tasks inspired by the Spaceport America Cup competition, requiring optimization for target altitude, structural integrity, landing precision, and cost efficiency—challenges that mirror real-world engineering constraints.

Our findings reveal that while state-of-the-art LLMs demonstrate impressive baseline knowledge of engineering principles, they struggle to iteratively improve designs through simulation feedback — a critical capability for real-world engineering tasks. Even after numerous iterations, these models plateau below human performance, indicating limitations in their ability to fully optimize complex multi-parameter designs.

By implementing reinforcement learning (RL) approaches, we demonstrate that these limitations can be overcome, enabling LLMs to achieve performance that surpasses human experts. Our RL-trained models achieved precision landings within 12 meters of targets and consistently outperformed human designs across multiple metrics, despite using a relatively modest 7B parameter model architecture.

The results point to a future where LLMs could transform engineering practice across numerous disciplines once effective interfaces with domain-specific tools are established. This work offers a glimpse of engineering's future, where LLMs serve not merely as information retrieval systems but as creative problem-solvers capable of generating and optimizing novel designs beyond human capabilities.

\paragraph{Key Contributions}
\begin{itemize}
\item RocketBench: A benchmark for evaluating LLMs' rocket design capabilities
\item Analysis of current LLMs strengths in engineering domains
\item Demonstration that RL-trained LLMs significantly outperform both larger foundation models and human experts.
\end{itemize}

\section{Related Work}
\paragraph{RL for Engineering Design}
Reinforcement learning has demonstrated significant efficacy in advancing engineering and scientific capabilities. AlphaFold transformed protein structure prediction\citep{jumper_highly_2021}, while AlphaTensor \citep{noauthor_discovering_2025} discovered novel matrix multiplication algorithms with superior computational efficiency (Fawzi et al., 2022). Similarly, AlphaDev\citep{noauthor_alphadev_2025} optimized low-level sorting algorithms beyond human-designed solutions, and Alpha Chip\citep{noauthor_how_2025} enhanced semiconductor design through systematic exploration of parameter spaces. These systems primarily leverage traditional RL methodologies with domain-specific reward structures. Our approach diverges by employing language models as the base policy, leveraging their baseline engineering knowledge and physical reasoning capabilities. This integration of structured domain priors with reinforcement learning enables more sample-efficient optimization while maintaining domain flexibility, addressing a fundamental limitation of previous approaches that require extensive task-specific engineering.

\section{Methodology}
\subsection{Simulation Env}
Our simulations were built on RocketPy, a high-fidelity trajectory simulation library for high-power rocketry. RocketPy provides a complete 6 degrees of freedom (6-DOF) simulation framework that accounts for variable mass effects, aerodynamic forces, and parachute descent phases with exceptional accuracy. The library has been validated against real-world flight data, demonstrating relative errors of less than 2\% for apogee predictions across multiple documented test flights\citep{RocketPy}.

For our research, we enhanced RocketPy with custom design rule checks (DRCs) and timeout mechanisms to address common failure modes observed during initial testing. DRCs included basic constraints including verification that body diameter exceeded motor diameter and that body length was greater than motor length to ensure proper component integration. These additions prevent simulation failures caused by physically impossible configurations and terminate excessively long computations that often result from unrealistic rocket parameters.

To enable LLMs to interface with the simulation environment, we developed a JSON interface that abstracts the complexity of RocketPy into a structured configuration format. Through this interface, models can specify all critical rocket parameters including motor selection, dimensions, materials, and launch conditions. See \autoref{Simulation_Vars} for full list.

Both motors and materials were limited to pre-curated lists of commercially available options to ensure physical realism and manufacturability. For motors, we selected a range of commercial solid rocket motors commonly used in high-power rocketry, varying in total impulse, burn duration, and thrust profiles. LLMs were passed detailed specifications for physical dimensions, propellant mass, and thrust information derived from manufacturer data. 

We also implemented a material stress simulation module that evaluates structural integrity throughout the flight profile. This module identifies the point of maximum stress during the mission and determines whether the rocket maintains structural integrity, adding a critical dimension of realism to the performance evaluation.

The simulation incorporated an economic model to introduce realistic cost constraints into the design process. Total costs were computed as a function of motor selection and material volume, with standardized per-unit prices assigned to each material type based on market averages. Motor costs were assigned fixed values reflecting their relative price differences while maintaining consistent cost pressure across designs. This economic dimension compelled LLMs to navigate performance-cost tradeoffs, simulating the budget optimization challenges inherent in real-world engineering projects. By incorporating this cost function into the reward signal, we ensured that design solutions reflected not only technical performance but also resource efficiency—a critical constraint in practical rocket engineering.

\begin{table}[htbp]
\centering
\caption{Configurable Rocket Parameters}
\begin{tabular}{lll}
\toprule
\textbf{Component} & \textbf{Parameter} & \textbf{Description} \\
\midrule
\multirow{1}{*}{Motor} & Motor Choice & Selection from available commercial motors \\
\midrule
\multirow{4}{*}{Rocket Body} & Radius & Body radius in meters (must exceed motor radius) \\
& Length & Body length in meters \\
& Material & Selection from available materials \\
& Thickness & Wall thickness in meters \\
\midrule
\multirow{3}{*}{Nose Cone} & Shape & Conical, ogive, elliptical, tangent, von karman, etc. \\
& Length & Nose cone length in meters \\
& Material & Selection from available materials \\
\midrule
\multirow{6}{*}{Fins} & Number & Number of fins around the rocket body \\
& Root Chord & Length of fin base in meters \\
& Tip Chord & Length of fin tip in meters \\
& Span & Fin height in meters \\
& Cant Angle & Fin angle in degrees \\
& Material & Selection from available materials \\
& Thickness & Fin thickness in meters \\
\midrule
\multirow{4}{*}{Tail} & Length & Tail section length in meters \\
& Top Radius & Upper radius of tail section in meters \\
& Bottom Radius & Lower radius of tail section in meters \\
& Material & Selection from available materials \\
\midrule
\multirow{6}{*}{Parachutes} & Main CD×S & Main parachute drag area \\
& Main Trigger & Deployment trigger condition \\
& Drogue CD×S & Drogue parachute drag area \\
& Drogue Trigger & Deployment trigger condition \\

\midrule
\multirow{3}{*}{Launch} & Rail Length & Launch rail length in meters \\
& Inclination & Launch angle from horizontal in degrees \\
& Heading & Compass heading in degrees \\
\midrule
\multirow{2}{*}{Payload} & Mass & Payload mass in kilograms \\
& Position & Relative position from rocket center in meters \\
\bottomrule
\end{tabular}
\label{Simulation_Vars}
\end{table}

\subsection{Competition Tasks}

We developed two increasingly complex tasks to evaluate model performance across a spectrum of engineering challenges. While the weighting and selection criteria for each scoring component were chosen heuristically, they were designed to approximate the evaluation frameworks used in actual rocket competitions.

\subsubsection{Target Altitude Challenge}

The first task was inspired by the 10,000-foot category of the Spaceport America Cup, an international collegiate rocket engineering competition. In this event, teams must design rockets that reach as close as possible to the target altitude while ensuring safe recovery and operational integrity. 

Our reward function was designed to capture the essential performance metrics of this competition, focusing on flight performance aspects that could be simulated. The reward balanced multiple objectives with the following components:

\begin{itemize}
    \item \textbf{Altitude Accuracy (50\%)}: A linear reward based on the percentage difference between the achieved apogee and the target altitude:
    \begin{equation}
        \text{percent\_difference} = \frac{|\text{apogee} - \text{target\_apogee}|}{\text{target\_apogee}}
    \end{equation}
    \begin{equation}
        \text{distance\_reward} = \max(0, 1.0 - \text{percent\_difference})
    \end{equation}

    \item \textbf{Structural Integrity (10\%)}: A binary component that awarded points only if the rocket maintained structural integrity throughout the flight:
    \begin{equation}
        \text{structural\_failure\_reward} = 
        \begin{cases}
            0 & \text{if structural failure occurred} \\
            1 & \text{otherwise}
        \end{cases}
    \end{equation}

    \item \textbf{Horizontal Drift Control (10\%)}: Models received higher scores for minimizing the horizontal displacement of landing position:
    \begin{equation}
        \text{max\_horz\_distance} = \text{target\_apogee} \times 0.3
    \end{equation}
    \begin{equation}
        \text{horz\_distance\_reward} = \max(0, 1 - \frac{\text{horizontal\_distance}}{\text{max\_horz\_distance}})
    \end{equation}

    \item \textbf{Cost Efficiency (15)}: A linear reward calculated based on the total cost of materials and components:
    \begin{equation}
        \text{cost\_reward} = \max(0, 1 - \frac{\text{total\_cost}}{1000})
    \end{equation}

    \item \textbf{Landing Safety (15\%)}: Models were rewarded for designs with safe descent velocities. Impact velocity in m/s:
    \begin{equation}
        \text{impact\_reward} = \max(0, 1 - \frac{\text{impact\_velocity}}{25})
    \end{equation}
\end{itemize}

The total reward was calculated as a weighted sum of these components:
\begin{align}
\boxed{
\begin{aligned}
R_{\text{total}} = 0.5 \cdot R_{\text{altitude}} + 0.1 \cdot R_{\text{structural}} + 0.1 \cdot R_{\text{drift}} + 0.15 \cdot R_{\text{cost}} + 0.15 \cdot R_{\text{landing}}
\end{aligned}
}
\end{align}

\subsubsection{Precision Landing Challenge}

The Second and most complex task shifted focus from altitude targeting to precision landing. Models were tasked with designing rockets capable of landing as close as possible to a target location 5.65 kilometers from the launch site (4000m horizontally and 4000m vertically) while maintaining cost efficiency and safety. The reward function for this challenge was structured as follows:

\begin{itemize}
    \item \textbf{Landing Accuracy (75\%)}: The primary component based on the distance between the actual landing location and the target coordinates:
    \begin{equation}
        \text{landing\_error} = \sqrt{(\text{landing\_x} - \text{target\_x})^2 + (\text{landing\_y} - \text{target\_y})^2}
    \end{equation}
    \begin{equation}
        \text{target\_distance} = \sqrt{\text{target\_x}^2 + \text{target\_y}^2}
    \end{equation}
    \begin{equation}
        \text{landing\_error\_percent} = \frac{\text{landing\_error}}{\text{target\_distance}}
    \end{equation}
    \begin{equation}
        \text{landing\_reward} = \max(0, 1.0 - \text{landing\_error\_percent})
    \end{equation}

    \item \textbf{Structural Integrity (15\%)}: A binary component that awarded points only if the rocket maintained structural integrity throughout the flight:
    \begin{equation}
        \text{structural\_failure\_reward} = 
        \begin{cases}
            0 & \text{if structural failure occurred} \\
            1 & \text{otherwise}
        \end{cases}
    \end{equation}

    \item \textbf{Cost Efficiency (5\%)}: A linear reward calculated based on the total cost of materials and components:
    \begin{equation}
        \text{cost\_reward} = \max(0, 1 - \frac{\text{total\_cost}}{1000})
    \end{equation}

    \item \textbf{Landing Safety (5\%)}: Models were rewarded for designs with safe descent velocities. Impact velocity in m/s:
    \begin{equation}
        \text{impact\_reward} = \max(0, 1 - \frac{\text{impact\_velocity}}{25})
    \end{equation}
\end{itemize}

The total reward was calculated as a weighted sum of these components:
\begin{align}
\boxed{
\begin{aligned}
R_{\text{total}} = 0.75 \cdot R_{\text{landing}} + 0.05 \cdot R_{\text{structural}} + 0.05 \cdot R_{\text{cost}} + 0.05 \cdot R_{\text{safety}}
\end{aligned}
}\end{align}

This challenge represented a significant increase in complexity, as it required models to reason about the entire flight trajectory, including both the powered ascent phase and the descent under parachutes. Success demanded sophisticated understanding of parachute deployment timing, wind drift compensation, and optimal flight path planning.

\subsection{RL Dynamics}
To explore the potential of reinforcement learning on LLMs for engineering tasks, we implemented Group Relative Policy Optimization (GRPO) using Qwen 2.5 7B model. Training was conducted with a batch size of 64 across all experiments.

For each training instance, the model received a standardized prompt containing the complete environmental specifications(wind conditions, target altitude or landing coordinates), available materials and engines, and the exact code implementation of the reward function. To establish baseline understanding, we provided a single example design in the expected output format. This ensured the model understood the task structure while minimizing prior solution bias. See appendix for full prompt

\subsection{Iterative Prompting Protocol}
For our comparative analysis of foundation models' iterative capabilities, we implemented a structured prompting protocol. Initial prompts contained identical information to the RL setup: environmental conditions, available components, and reward calculation methods.

In subsequent iterations, we augmented the prompt with the model's full output including reasoning for the previous solution and the comprehensive performance metrics. We include all previous solutions not just previous attempt. For the Target Altitude Challenge, feedback included maximum apogee achieved, structural integrity status, and final cost. For the Precision Landing Challenge, we substituted landing position for apogee information. This approach enabled models to learn from previous design attempts while maintaining consistent evaluation conditions.

\subsection{Human Baseline}
To establish a comparative baseline, we recruited an individual with multiple years of experience in university-level rocketry competitions similar to those our benchmarks are based off. This participant completed identical design tasks with the same interface, components, and feedback mechanisms used by the models. 

While this approach provides a meaningful comparison point, we note that this represents a single expert rather than peak human performance or professional expertise in the aerospace industry. The baseline serves as a practical reference point for assessing model performance against experienced human engineering capabilities under identical constraints.

\section{Results}
\subsection{Rocket Bench}

\subsubsection{Target Altitude Challenge}

\pgfplotsset{compat=1.18}

\definecolor{gpt4o}{RGB}{82, 177, 255}     
\definecolor{claude}{RGB}{156, 39, 176}     
\definecolor{o1}{RGB}{0, 200, 83}           
\definecolor{o1mini}{RGB}{255, 152, 0}       
\definecolor{humans}{RGB}{244, 67, 54}       
\definecolor{avgblue}{HTML}{2E74B5}          
\definecolor{bestred}{HTML}{C00000}          
\definecolor{gridcolor}{HTML}{E0E0E0}
\definecolor{textcolor}{HTML}{333333}

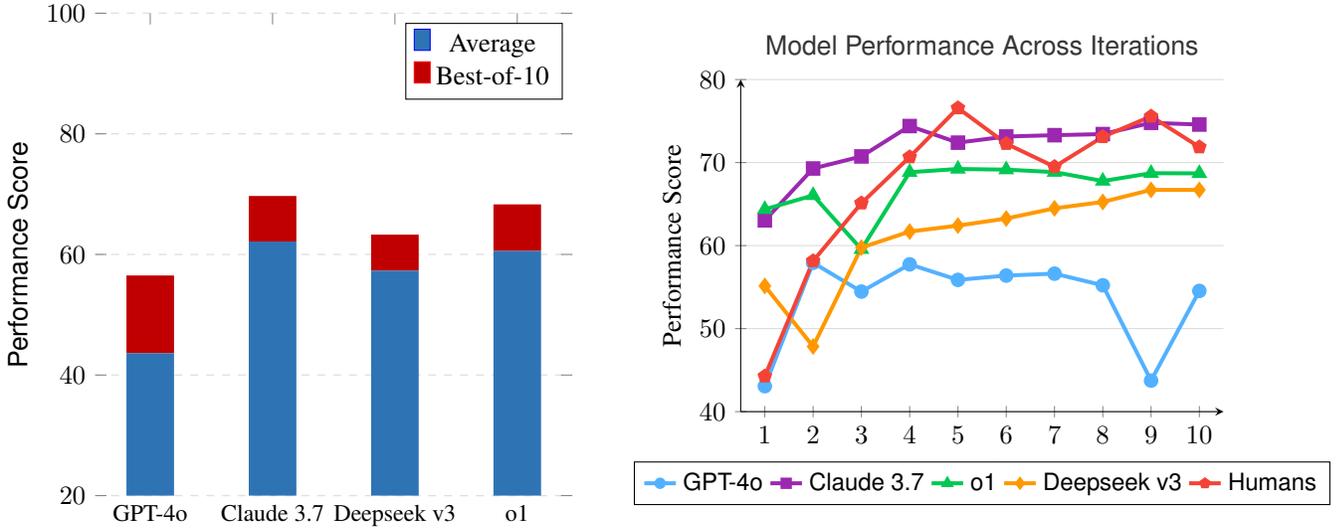
\begin{figure}[htbp]
  \centering
  \begin{minipage}{0.48\textwidth}
    \centering
    \begin{tikzpicture}
      \begin{axis}[
        ybar stacked,
        bar width=18pt,
        width=\textwidth,
        height=8cm,
        enlarge x limits=0.15,
        symbolic x coords={GPT-4o, Claude 3.7, Deepseek v3, o1},
        xtick=data,
        xticklabel style={align=center,font=\small},
        ylabel={\sffamily Performance Score},
        ymin=20,
        ymax=100,
        axis line style={draw=none},
        ymajorgrids=true,
        grid style={gridcolor, dashed, line width=0.5pt},
        x axis line style={gridcolor, line width=0.5pt},
        clip=false,
        xtick align=inside,
    ]

      \addplot+[fill=avgblue, draw=none] coordinates {
          (GPT-4o,43.62)
          (Claude 3.7,62.14)
          (Deepseek v3,57.3)
          (o1,60.57)
      };

      \addplot+[fill=bestred, draw=none] coordinates {
          (GPT-4o,12.89)
          (Claude 3.7, 7.55)
          (Deepseek v3,5.98)
          (o1,7.73)
      };

      \legend{Average, Best‐of‐10}

    \end{axis}
    \end{tikzpicture}
  \end{minipage}%
  \hfill
  \begin{minipage}{0.48\textwidth}
    \centering
    \begin{tikzpicture}
      \begin{axis}[
          width=8cm,  
          height=6cm, 
          title={Model Performance Across Iterations},
          title style={font=\sffamily, text=textcolor}, 
          xlabel={Iteration},
          ylabel={Performance Score},
          xmin=0.5, xmax=10.5,
          ymin=40, ymax=80,
          xtick={1,2,3,4,5,6,7,8,9,10},
          ytick={20,30,40,50,60,70,80},
          ymajorgrids=true,
          grid style={line width=.1pt, draw=gray!10},
          major grid style={line width=.2pt, draw=gray!30},
          axis lines=left,
          every axis plot/.append style={line width=1.5pt, mark size=2pt},
          legend style={
              font=\sffamily\small,
              legend image post style={scale=0.7},
              at={(0.5,-0.15)},
              anchor=north,
              legend columns=5,
          },
      ]
      
      \addplot[color=gpt4o, mark=*, mark options={fill=gpt4o}] coordinates {
          (1, 43.07)
          (2, 57.93)
          (3, 54.46)
          (4, 57.74)
          (5, 55.87)
          (6, 56.39)
          (7, 56.62)
          (8, 55.22)
          (9, 43.74)
          (10, 54.54)
      };
      
      \addplot[color=claude, mark=square*, mark options={fill=claude}] coordinates {
          (1, 63.04)
          (2, 69.29)
          (3, 70.74)
          (4, 74.4)
          (5, 72.4)
          (6, 73.14)
          (7, 73.3)
          (8, 73.43)
          (9, 74.79)
          (10, 74.58)
      };
      
      \addplot[color=o1, mark=triangle*, mark options={fill=o1}] coordinates {
          (1, 64.37)
          (2, 66.05)
          (3, 59.56)
          (4, 68.84)
          (5, 69.23)
          (6, 69.15)
          (7, 68.85)
          (8, 67.78)
          (9, 68.72)
          (10, 68.7)
      };
      
      \addplot[color=o1mini, mark=diamond*, mark options={fill=o1mini}] coordinates {
            (1, 55.14)
            (2, 47.85)
            (3, 59.76)
            (4, 61.69)
            (5, 62.41)
            (6, 63.26)
            (7, 64.49)
            (8, 65.26)
            (9, 66.71)
            (10,66.71)
        };

      \addplot[color=humans, mark=pentagon*, mark options={fill=humans}] coordinates {
          (1, 44.27)
          (2, 58.15)
          (3, 65.11)
          (4, 70.67)
          (5, 76.57)
          (6, 72.28)
          (7, 69.5)
          (8, 73.1)
          (9, 75.57)
          (10, 71.87)
      };
      
      \legend{GPT-4o, Claude 3.7, o1, Deepseek v3, Humans}
      \end{axis}
    \end{tikzpicture}
  \end{minipage}

  \caption{Performance across Target Altitude Challenge. Left shows average performance across 10 runs right shows sample progress over time (Best over 3 runs for LLMs)}
  \label{fig:consistent_charts}
\end{figure}

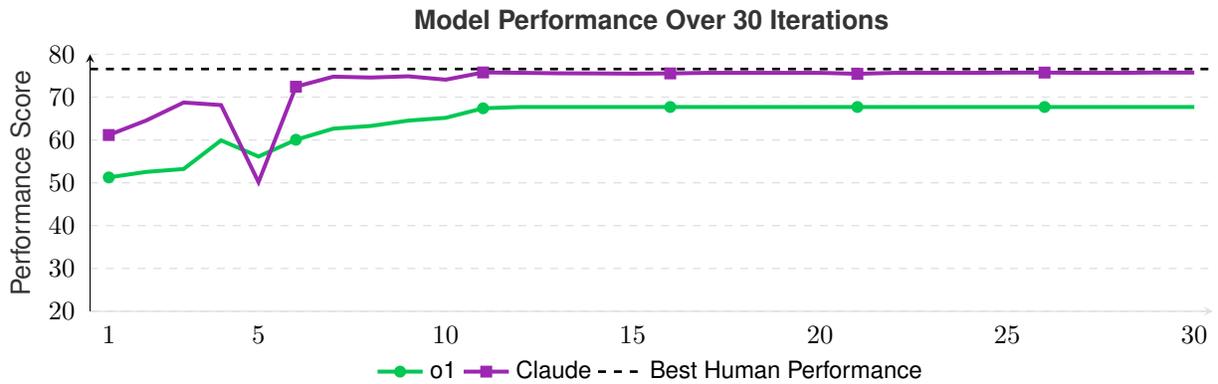
\begin{figure}
\centering
\begin{tikzpicture}
\begin{axis}[
    width=\textwidth,
    height=5cm,
    title={Model Performance Over 30 Iterations},
    title style={font=\sffamily\bfseries, text=textcolor},
    xlabel={Iteration},
    xlabel style={font=\sffamily, text=textcolor},
    ylabel={Performance Score},
    ylabel style={font=\sffamily, text=textcolor},
    xmin=0.5, xmax=30.5,
    ymin=20, ymax=80,
    xtick={1,5,10,15,20,25,30},
    ytick={20,30,40,50,60,70,80},
    ymajorgrids=true,
    grid style={gridcolor, dashed, line width=0.5pt},
    axis lines=left,
    tick style={draw=none},
    x axis line style={gridcolor, line width=0.5pt},
    legend style={
        at={(0.5,-0.15)},
        anchor=north,
        legend columns=3,
        font=\sffamily\small,
        draw=none,
        fill=white
    },
    legend cell align={left},
    every axis plot/.append style={
      line width=1.5pt, 
      mark size=1.5pt, 
      mark repeat=5
    },
]

\addplot[color=o1, mark=*, mark options={fill=o1}] coordinates {
    (1,51.26)
    (2,52.54)
    (3,53.22)
    (4,59.91)
    (5,56.13)
    (6,60.05)
    (7,62.65)
    (8,63.27)
    (9,64.52)
    (10,65.17)
    (11,67.4)
    (12,67.7)
    (13,67.7)
    (14,67.7)
    (15,67.7)
    (16,67.7)
    (17,67.7)
    (18,67.7)
    (19,67.7)
    (20,67.7)
    (21,67.7)
    (22,67.7)
    (23,67.7)
    (24,67.7)
    (25,67.7)
    (26,67.7)
    (27,67.7)
    (28,67.7)
    (29,67.7)
    (30,67.7)
};

\addplot[color=claude, mark=square*, mark options={fill=claude}] coordinates {
    (1, 61.16)
    (2, 64.55)
    (3, 68.75)
    (4, 68.15)
    (5, 50.16)
    (6, 72.44)
    (7, 74.79)
    (8, 74.58)
    (9, 74.87)
    (10, 74.08)
    (11, 75.77)
    (12, 75.69)
    (13, 75.57)
    (14, 75.54)
    (15, 75.49)
    (16, 75.54)
    (17, 75.68)
    (18, 75.68)
    (19, 75.68)
    (20, 75.68)
    (21, 75.45)
    (22, 75.68)
    (23, 75.68)
    (24, 75.68)
    (25, 75.73)
    (26, 75.74)
    (27, 75.68)
    (28, 75.68)
    (29, 75.75)
    (30, 75.73)
};

\addplot [domain=0.5:30.5, dashed, color=black, line width=1pt] {76.57};
\addlegendentry{Best Human Performance}

\legend{o1, Claude, Best Human Performance}
\end{axis}
\end{tikzpicture}
\caption{Performance comparison of Claude vs. o1 over 30 iterations with best human performance indicated.}
\label{fig:30iteration}
\end{figure}

The Target Altitude Challenge results revealed interesting performance patterns across tested models. Claude 3.7\citep{noauthor_claude_nodate} achieved the highest average performance (62.14), with Deepseek v3\citep{deepseekv3} (57.36) and o1\citep{o1} (60.57) following closely behind. The relative similarity between these top three models' scores suggests comparable engineering reasoning capabilities for this task. 

All models demonstrated strong initial designs, generally outperforming early human attempts. This finding is particularly noteworthy as it indicates that LLMs possess effective baseline engineering intuition even without specialized training. The models' ability to immediately generate viable rocket designs suggests they have internalized fundamental physics and engineering principles during their general training.

We next explored how models performed at iterating and improving on their designs. \autoref{fig:consistent_charts} (right panel) reveals distinct patterns in how models improved their designs through iteration. All models demonstrated the ability to learn from feedback and improve their designs, but at significantly different rates. Claude 3.7 showed the strongest overall progression, starting at 63.04 and reaching 74.79 by iteration 9. This consistent upward trajectory suggests effective learning from simulation feedback. o1 began with a similarly strong baseline score of 64.37 and stabilized around 68-69 after several iterations, showing good but less dramatic improvement over time. Deepseek v3 demonstrated steady improvement from a moderate starting score of 52.14 to approximately 66.71 by iteration 10. GPT-4o \citep{gpt4o} showed inconsistent performance between 43-57 without clear convergence, indicating difficulty in systematically incorporating feedback into design improvements.

Human experts began at a comparable performance level (44.27) but improved more rapidly and effectively, reaching a maximum score of 76.57 by iteration 5 - the highest score achieved in our study. This demonstrates humans' superior ability to iterate on designs based on simulation feedback. Despite strong baseline capabilities and consistent improvement trends, all LLMs ultimately fell short of human performance in iterative design optimization.

We next investigated whether additional iterations would enable models to surpass human performance. Our 30-iteration experiment with Claude 3.7 and o1 (\autoref{fig:30iteration} yielded particularly insightful results about the models' improvement limitations. Both models plateaued below the human maximum despite extensive iteration opportunities. Claude 3.7 stabilized around 75.73, coming close to but never surpassing the human record of 76.57, while o1 plateaued at a lower 67.7. This persistent gap indicates that neither model could match the human level performance despite having triple the iteration cycles compared to the baseline experiment.

We also evaluated "best-of-10" sampling alongside our primary methods, observing significant performance gains across models. The improvements varied by model: GPT-4o showed the largest relative improvement, gaining 12.89 points to reach 56.51; o1 increased from 60.57 to 68.3; Claude 3.7 rose from 62.14 to 69.69; and Deepseek v3 improved from 57.3 to 63.28. Despite these gains, all models achieved even higher peak scores through iterative prompting compared to best-of-10 sampling.

\subsubsection{Precision Landing Challenge}

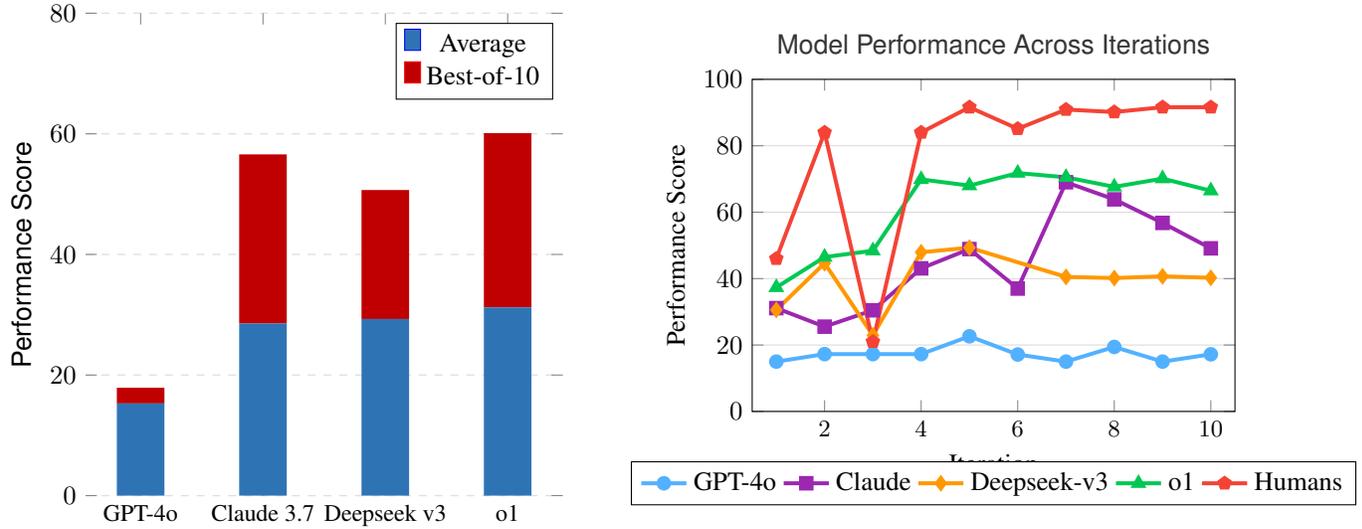
\begin{figure}[htbp]
  \centering
  \begin{minipage}{0.48\textwidth}
    \centering
    \begin{tikzpicture}
      \begin{axis}[
        ybar stacked,
        bar width=18pt,
        width=\textwidth,
        height=8cm,
        enlarge x limits=0.15,
        symbolic x coords={GPT-4o, Claude 3.7, Deepseek v3, o1},
        xtick=data,
        xticklabel style={align=center,font=\small},
        ylabel={\sffamily Performance Score},
        ymin=0,
        ymax=80,
        axis line style={draw=none},
        ymajorgrids=true,
        grid style={gridcolor, dashed, line width=0.5pt},
        x axis line style={gridcolor, line width=0.5pt},
        clip=false,
        xtick align=inside,
    ]

      \addplot+[fill=avgblue, draw=none] coordinates {
          (GPT-4o,15.29)
          (Claude 3.7,28.54)
          (Deepseek v3,29.29)
          (o1,31.21)
      };

      \addplot+[fill=bestred, draw=none] coordinates {
          (GPT-4o,2.6)
          (Claude 3.7, 28.05)
          (Deepseek v3,21.39)
          (o1,28.9)
      };

      \legend{Average, Best‐of‐10}

    \end{axis}
    \end{tikzpicture}
  \end{minipage}%
  \hfill
  \begin{minipage}{0.48\textwidth}
    \centering
    \begin{tikzpicture}
      \begin{axis}[
          width=8cm,  
          height=6cm, 
          title={Model Performance Across Iterations},
          title style={font=\sffamily, text=textcolor},
          xlabel={Iteration},
          ylabel={Performance Score},
          xmin=0.5, xmax=10.5,
          ymin=0, ymax=100,
          ymajorgrids=true,
          grid style={line width=.1pt, draw=gray!10},
          major grid style={line width=.2pt, draw=gray!30},
          xticklabel style={font=\small},
          every axis plot/.append style={line width=1.5pt, mark size=2pt},
          legend style={
              at={(0.5,-0.15)},
              anchor=north,
              legend columns=5,
          },
      ]
      
      \addplot[color=gpt4o, mark=*, mark options={fill=gpt4o}] coordinates {
          (1,15)
          (2,17.26)
          (3,17.26)
          (4,17.26)
          (5,22.63)
          (6,17.14)
          (7,15)
          (8,19.42)
          (9,15)
          (10,17.2)
      };
      
      \addplot[color=claude, mark=square*, mark options={fill=claude}] coordinates {
          (1,31.11)
          (2,25.53)
          (3,30.46)
          (4,43.08)
          (5,48.89)
          (6,37)
          (7,69.01)
          (8,63.86)
          (9,56.77)
          (10,49.11)
      };
      
      \addplot[color=o1mini, mark=diamond*, mark options={fill=o1mini}] coordinates {
            (1, 30.57)
            (2, 44.64)
            (3, 22.91)
            (4, 47.92)
            (5, 49.32)
            (7, 40.50)
            (8, 40.16)
            (9, 40.68)
            (10, 40.24)
        };

      \addplot[color=o1, mark=triangle*, mark options={fill=o1}] coordinates {
          (1,37.34)
          (2,46.49)
          (3,48.39)
          (4,69.88)
          (5,68)
          (6,71.78)
          (7,70.51)
          (8,67.63)
          (9,70.11)
          (10,66.46)
      };
      
      \addplot[color=humans, mark=pentagon*, mark options={fill=humans}] coordinates {
          (1,46)
          (2,84)
          (3,21)
          (4,84)
          (5,91.6)
          (6,85.12)
          (7,90.9)
          (8,90.15)
          (9,91.6)
          (10,91.6)
      };
      
      \legend{GPT-4o, Claude, Deepseek-v3, o1, Humans}
      \end{axis}
    \end{tikzpicture}
  \end{minipage}

  \caption{Performance across Precision Landing Challenge. Left shows average and best performance across 10 runs right shows sample progress over time (Best over 3 runs for LLMs)}
  \label{fig:precision_landing}
\end{figure}

The Precision Landing Challenge results seeked to explore how LLMs handled more complex environments. ~\autoref{fig:precision_landing} (left panel) showed O1 achieved the highest average performance (31.21), with Deepseek-v3 (29.29) and Claude 3.7 (28.54) following closely. GPT-4o scored substantially lower (15.29), indicating particular difficulty with this more complex task.

Unlike the Altitude Challenge, all models demonstrated lower initial performance relative to human experts in this task, suggesting that the precision landing requirements introduce complexity that current LLMs struggle to address effectively.

\autoref{fig:precision_landing} (right panel) reveals more pronounced differences in iterative capabilities. Human experts demonstrated rapid improvement, beginning at 46 and rapidly improving to scores above 90 by iteration 5. O1 showed the strongest model performance, reaching scores around 70 in iterations 4-9, but with inconsistent patterns. Claude 3.7 demonstrated highly variable performance, with peaks near 69 but substantial fluctuations. Deepseek-v3 showed moderate improvement but plateaued around 40-45. GPT-4o struggled significantly, never exceeding 23 points. We note high variance in iteration scores between runs but general performance trends remained consistent throughout all trials.

The performance gap between humans and even the best-performing models is substantially wider than in the Altitude Challenge. While the best human score reached 91.60, the top model score (o1) peaked at 71.78, showing a 19.82-point differential compared to the 2-point gap in the previous task.

This pronounced performance difference highlights how increasing task complexity disproportionately affects LLM capabilities. Unlike the Target Altitude Challenge where models approached human performance, the Precision Landing Challenge reveals more substantial limitations in models' ability to reason through complex engineering environments

We also tested "best-of-10" sampling alongside iterative prompting, finding substantial performance improvements: o1 increased from 31.21 to 60.11, Claude 3.7 from 28.54 to 56.59, and Deepseek v3 gained 21.39 points to reach 50.68. GPT 4o struggled to make any substantial improvements. These improvements again fell short of those achieved through iterative prompting, where models reached higher peak scores, demonstrating that LLMs genuinely learn from feedback rather than merely benefiting from sampling variation. 

Our results across both challenges reveal a consistent pattern: LLMs demonstrate impressive baseline engineering knowledge but significant limitations in iterative design capabilities. In the Target Altitude Challenge, models produced initial designs comparable to human experts, with some models approaching (but never surpassing) human performance after extensive iteration. However, the more complex Precision Landing Challenge exposed a much wider performance gap, with even the best models scoring 19.82 points below human experts.

This pattern suggests that while current LLMs have internalized substantial engineering principles from their training, they lack the strategic iteration abilities that human experts employ when refining designs. These findings indicate that while LLMs show promise as engineering tools for generating initial designs and baseline solutions, they currently cannot match human experts' ability to iteratively refine complex engineering systems through feedback-driven optimization.

\definecolor{deepblue}{RGB}{31, 119, 180}     

\definecolor{deepblue}{RGB}{0,0,139}
\definecolor{textblue}{RGB}{0,0,139}
\definecolor{textcolor}{RGB}{0,0,0} 
\definecolor{gridcolor}{RGB}{200,200,200}

\pgfplotsset{
  same axis style/.style={
    width=\textwidth,
    height=5cm,
    grid style={gridcolor,dashed,line width=0.5pt},
    tick style={draw=none},
    axis lines=left,
    x axis line style={gridcolor, line width=0.5pt},
    axis background/.style={fill=white},
    every axis plot/.append style={line width=1.5pt},
    title style={font=\sffamily\bfseries, },
    xlabel style={font=\sffamily, },
    ylabel style={font=\sffamily, },
  }
}

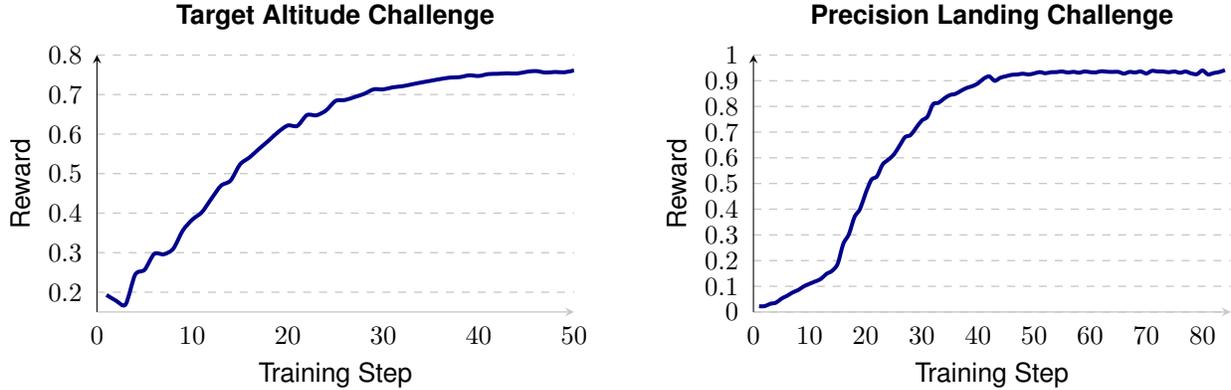
\begin{figure}[htb]
  \centering
  \begin{minipage}[b]{0.48\textwidth}
    \centering
    \begin{tikzpicture}
      \begin{axis}[
        same axis style,
        title={Target Altitude Challenge},
        xlabel={Training Step},
        ylabel={Reward},
        xmin=0, xmax=50,
        ymin=0.15, ymax=0.8,
        xtick={0,10,20,30,40,50},
        ytick={0.2,0.3,0.4,0.5,0.6,0.7,0.8},
        ymajorgrids=true,grid style={gridcolor, dashed, line width=0.5pt},
      ]
      \addplot[color=deepblue, mark=none, smooth] coordinates {
          (1, 0.19312062862)
          (2, 0.17850108443)
          (3, 0.16903597124)
          (4, 0.24417218575)
          (5, 0.25697344546)
          (6, 0.29692268377)
          (7, 0.29593598848)
          (8, 0.31006777299)
          (9, 0.3562153876)
          (10, 0.3838055432)
          (11, 0.4025335312)
          (12, 0.4368537962)
          (13, 0.4695712626)
          (14, 0.4823873043)
          (15, 0.5235486627)
          (16, 0.5412806273)
          (17, 0.5629762411)
          (18, 0.5836820602)
          (19, 0.6055434942)
          (20, 0.6218647361)
          (21, 0.620785594)
          (22, 0.6480969787)
          (23, 0.6475751996)
          (24, 0.659719348)
          (25, 0.6838907003)
          (26, 0.6862820983)
          (27, 0.6937907338)
          (28, 0.7018692493)
          (29, 0.713260293)
          (30, 0.713101089)
          (31, 0.7183753252)
          (32, 0.7211125493)
          (33, 0.7259053588)
          (34, 0.7307115197)
          (35, 0.734999001)
          (36, 0.7393943667)
          (37, 0.7429887056)
          (38, 0.7440939546)
          (39, 0.7487320304)
          (40, 0.7470393777)
          (41, 0.7518469095)
          (42, 0.7525888681)
          (43, 0.7537796497)
          (44, 0.75327)
          (45, 0.7574740648)
          (46, 0.7595540881)
          (47, 0.7557290792)
          (48, 0.7572389245)
          (49, 0.7564098835)
          (50, 0.7612124681)
      };
      \end{axis}
    \end{tikzpicture}
  \end{minipage}
  \hfill
  \begin{minipage}[b]{0.48\textwidth}
    \centering
    \begin{tikzpicture}
      \begin{axis}[
        same axis style,
        title={Precision Landing Challenge},
        xlabel={Training Step},
        ylabel={Reward},
        xmin=0, xmax=85,
        ymin=0, ymax=1.0,
        xtick={0,10,20,30,40,50,60,70,80},
        ytick={0,0.1,0.2,0.3,0.4,0.5,0.6,0.7,0.8,0.9,1.0},
        ymajorgrids=true,grid style={gridcolor, dashed, line width=0.5pt},
      ]
      \addplot[color=deepblue, mark=none, smooth] coordinates {
          (1, 0.022426880452)
          (2, 0.022522381993)
          (3, 0.03177893494)
          (4, 0.036010067465)
          (5, 0.051925625216)
          (6, 0.062939602437)
          (7, 0.076063151968)
          (8, 0.085676770219)
          (9, 0.09998094618)
          (10, 0.1097009885)
          (11, 0.1187804151)
          (12, 0.1284326422)
          (13, 0.1478950489)
          (14, 0.1599294055)
          (15, 0.1870409358)
          (16, 0.2646932709)
          (17, 0.3024593103)
          (18, 0.3693806338)
          (19, 0.4012997556)
          (20, 0.4633521843)
          (21, 0.5142459679)
          (22, 0.5279465723)
          (23, 0.5735294032)
          (24, 0.5924203205)
          (25, 0.6123809028)
          (26, 0.6459426689)
          (27, 0.6801860142)
          (28, 0.6885850239)
          (29, 0.7174548912)
          (30, 0.7450775909)
          (31, 0.7600340652)
          (32, 0.8083923864)
          (33, 0.8143405724)
          (34, 0.8311622548)
          (35, 0.8441992807)
          (36, 0.8488880801)
          (37, 0.8611078072)
          (38, 0.8720918703)
          (39, 0.8788429308)
          (40, 0.8896862078)
          (41, 0.907728653)
          (42, 0.9169090199)
          (43, 0.8997243166)
          (44, 0.9117851377)
          (45, 0.9175301313)
          (46, 0.9229427099)
          (47, 0.9242611051)
          (48, 0.9274011374)
          (49, 0.9243204117)
          (50, 0.9283468246)
          (51, 0.9336416245)
          (52, 0.9290891409)
          (53, 0.9330671549)
          (54, 0.9335659266)
          (55, 0.9360725403)
          (56, 0.931862545)
          (57, 0.9346519232)
          (58, 0.9312052846)
          (59, 0.9360755801)
          (60, 0.9333527803)
          (61, 0.9325658798)
          (62, 0.937077415)
          (63, 0.9349386215)
          (64, 0.9342687845)
          (65, 0.9345875502)
          (66, 0.9276450992)
          (67, 0.9342942357)
          (68, 0.9322514653)
          (69, 0.9359927297)
          (70, 0.9282167673)
          (71, 0.938509953)
          (72, 0.9359778285)
          (73, 0.9358018756)
          (74, 0.9330431938)
          (75, 0.9359293699)
          (76, 0.9310562134)
          (77, 0.9358732224)
          (78, 0.9284137607)
          (79, 0.9254682779)
          (80, 0.9398773432)
          (81, 0.9243745923)
          (82, 0.9300373912)
          (83, 0.9339650393)
          (84, 0.9413020134)
      };
      \end{axis}
    \end{tikzpicture}
  \end{minipage}
  \caption{Comparison of RL performance. Reward represents mean batch reward}
  \label{fig:comparison}
\end{figure}

\subsection{Reinforcement Learning}

\subsubsection{Target Apogee Challenge}
We next explored the effectiveness of applying RL on LLMs to engineering tasks. \autoref{fig:comparison} (left) shows the training progression for our RL with a Qwen-2.5 7B model. Over 50 training steps, the model demonstrated consistent improvement in mean batch reward from 0.19 to 0.76, indicating systematic optimization beyond what iterative prompting achieved.

The RL-trained model achieved a peak score of 79.98 on the Target Altitude Challenge, surpassing both the best human expert score (76.57) and all tested state-of-the-art LLMs. This performance differential is particularly notable given the model's significantly worse initial performance demonstrating that we do not need relatively high initial performance to make progress with RL.

\subsubsection{Precision Landing Challenge}

The Precision Landing Challenge results showed even more dramatic improvements when applying reinforcement learning. As illustrated in \autoref{fig:comparison} (right), the RL-trained 7B Qwen-2.5 model demonstrated exceptional progress over 84 training steps, with mean batch scores increasing from near-zero (0.02) to approximately 0.94.

Most significantly, the RL approach achieved a peak score of 95.6 on this challenge, substantially exceeding both human expert performance (91.6) and the best-performing foundation models (71.78 for o1). The trained model achieved remarkable precision, landing within just 12 meters of the target location—a feat that neither humans nor standard LLMs could match.

The dramatic performance improvements achieved through reinforcement learning across both challenges demonstrate the exceptional potential of RL-optimized LLMs for engineering applications. While iterative prompting of SoTA foundation models showed clear performance plateaus below human expert levels, RL training enabled consistent improvement beyond human performance thresholds. The ability of a relatively compact 7B parameter model to systematically outperform both larger foundation models and skilled human engineers indicates that RL optimization effectively addresses the core limitation we observed in standard LLMs: their difficulty in effectively altering designs given feedback. As engineering complexity increased from altitude targeting to precision landing, the RL advantage became more pronounced, suggesting this approach may be particularly valuable for more complex, multi-objective engineering optimization problems.


\begin{figure}[htbp]
    \centering
    \includegraphics[width=0.8\textwidth]{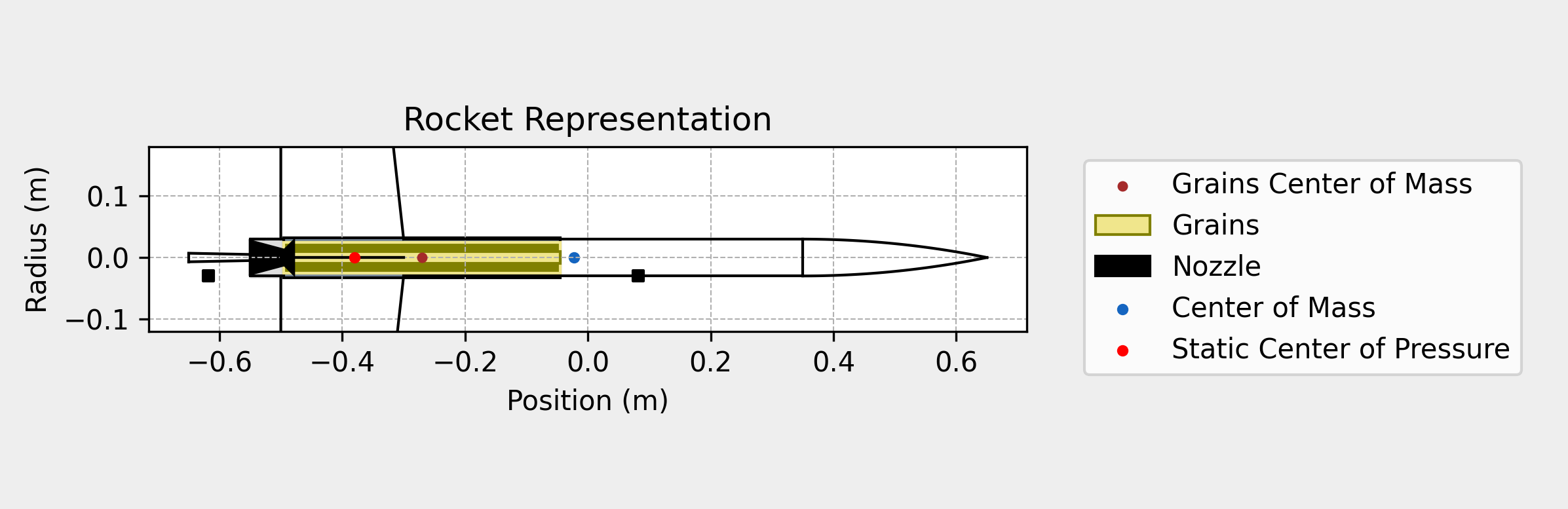}
    \caption{Best Rocket Design from RL on Precision Landing Challenge}
    \label{fig:rocket-performance}
\end{figure}






\section{Limitations}
While our simulation environment provides a reasonable proxy for evaluating LLMs' engineering capabilities, several limitations should be acknowledged. First, RocketPy, though high-fidelity, cannot account for all manufacturing complexities and environmental factors that physical rocket construction would entail. Second, we observed high variance in model performance, where small parameter adjustments sometimes led to disproportionately large differences in flight outcomes. This sensitivity reflects real engineering challenges but complicates consistent performance assessment. Third, our human baseline was limited to a single expert with ten attempts, matching the constraints imposed on the models.  While this provides a fair comparison, we acknowledge that human experts with additional iterations or collaborative environments might develop superior solutions. Despite these limitations, our results demonstrate that RL-enhanced LLMs can outperform individual human experts under controlled conditions, establishing a meaningful benchmark for future research.

\section{Discussion}

\subsection{RL vs Inference Time Compute}
Our results indicate that inference time compute scaling works across both simple and complex engineering tasks but current methods still plateau below human level performance. We observe that reasoning-focused models (o1) outperform foundation models (GPT-4o), suggesting that further scaling of the reasoning model paradigm could yield additional performance improvements.

Test-time reinforcement learning (TTRL) represents another promising axis for scaling compute to enhance performance. Beyond providing an incremental boost in specific domains, TTRL offers a methodology for extending performance boundaries in individual domains, potentially previewing future capabilities of large language models.




\subsection{Why use LLMs for RL?}
Despite impressive successes like AlphaGo and AlphaZero in game environments, reinforcement learning has seen limited application in engineering domains. One of the reasons for this is the cold-start problem in complex domains—traditional RL approaches require extensive random exploration before achieving meaningful performance, making them impractical for computationally expensive engineering simulations. Additionally, these systems typically need meticulously crafted environments with specialized reward functions, limiting their broad applicability.

LLMs offer a compelling solution to these challenges by leveraging their pre-trained knowledge of physics, engineering principles, and domain-specific concepts. This built-in prior knowledge enables them to begin optimization from reasonable starting points rather than random exploration. Our experiments demonstrate remarkable sample efficiency, with the RL-trained model requiring only approximately 3,000 simulation samples to exceed human-level performance on both tasks—orders of magnitude fewer than traditional RL approaches would require for problems of similar complexity.

This efficiency is particularly important given the long simulation times for complex engineering tasks, where each evaluation may require significant computational resources. The efficiency stems from the LLM's ability to make informed parameter adjustments based on underlying principles rather than blind trial-and-error. The combination of broad knowledge representation and targeted optimization through RL creates a powerful paradigm for accelerating engineering design processes that would be prohibitively expensive using either approach independently.

\subsection{Engineering Implication}
Our research demonstrates that given appropriate simulation environments, RL-trained LLMs can surpass human-level performance in complex engineering domains. Currently, two primary bottlenecks limit broader application: creating environments that interface effectively with LLMs and developing appropriate reward models. As models' agentic capabilities increase, we anticipate access to a wider range of possible RL environments. Similarly, improvements in models' self-verification abilities will enable setting more abstract goals, potentially significantly openning up the scope of possible RL tasks. If these challenges are addressed, we foresee rapid progress in engineering domains with strong simulation tools. The pattern mirrors what we've observed with diffusion models in image generation—where effective training methodologies led to explosive capability improvements that transformed creative industries within a remarkably short timeframe. The economic and scientific implications of this acceleration could be substantial across multiple engineering disciplines.

In the short term, we anticipate these systems will function as next-generation CAD tools for engineers, automatically exploring design spaces and suggesting optimized solutions. Just as traditional CAD software transformed engineering by automating drafting and visualization tasks, LLMs could automate aspects of design optimization while enabling human engineers to focus on higher-level innovation. This human-AI collaboration could significantly accelerate engineering progress across fields from aerospace to renewable energy and biomedical devices.

\subsection{Safety}
This research highlights broader security concerns as RL-enhanced LLMs could significantly lower barriers to developing dangerous technologies. The ability to rapidly optimize designs through simulation feedback potentially democratizes knowledge that was previously restricted through classification and specialized expertise requirements. 

These developments raise critical challenges for open source communities and AI regulation. With even current models demonstrating concerning capabilities through new inference scaling techniques, traditional regulatory approaches focused primarily on limiting training compute appear increasingly ineffective. Novel governance frameworks must address the emerging reality that test-time optimization and inference-focused methods can bypass conventional safeguards

\subsection{Conclusion}
Our research demonstrates that reinforcement learning applied to LLMs creates a powerful paradigm for engineering optimization that surpasses both foundation models and human expert performance. While state-of-the-art LLMs show strong baseline engineering knowledge, they consistently plateau below human capabilities when iteratively refining designs. In contrast, RL-trained models achieved performance breakthroughs in both target altitude and precision landing challenges using only a modest parameter architecture.

This approach addresses core limitations of both traditional RL (sample inefficiency) and foundation models (limited optimization capabilities), combining LLMs' domain knowledge with structured exploration. The results suggest a future where RL-optimized LLMs serve as force-multipliers across engineering disciplines, dramatically accelerating design cycles and enabling performance improvements previously unattainable. As interfaces between LLMs and simulation environments improve, we anticipate widespread adoption across aerospace, civil engineering, energy systems, and beyond—fundamentally transforming how complex engineering challenges are solved.

\appendix
\section{Sample Prompt}
\begin{lstlisting}
# Rocket Design Task

## Design Requirements

- **Target Apogee**: 3048.0 meters
- **Wind Conditions**: 5 m/s from E direction

## You are scored off the following
Distance to max apogee
Cost: Cheaper the rocket the better the score
Does it land safely (less than 5 m/s)
Does it not break
Horz distance: How far is it from the intial launch site

#Score func code shown below:
structural_failure = None  # Whether the rocket structure failed during flight
distance_reward = 1.0 - percent_difference
distance_reward = max(0, distance_reward)




# Structural failure reward
structural_failure_reward = 0 if structural_failure else 1

# Horizontal distance reward (linear version)
max_horz_distance = target_apogee * 0.3  # Scale factor
horz_distance_reward = max(0, 1 - horizontal_distance / max_horz_distance)

# Cost reward (linear version)
max_cost = 1000.0  # Base cost scale
cost_factor = total_cost / max_cost
cost_reward = 1.0 - cost_factor
cost_reward = max(0, cost_reward)  # Clamp to minimum of 0

# Impact velocity reward (linear version)
max_impact_velocity = 25  # m/s
impact_factor = abs(impact_velocity) / max_impact_velocity
impact_reward = 1.0 - impact_factor
impact_reward = max(0, impact_reward)  # Clamp to minimum of 0

# Add additional rewards with weights
reward = (distance_reward*0.5 +
            horz_distance_reward * 0.1 + 
            cost_reward * 0.15 + 
            impact_reward * 0.15 + 
            structural_failure_reward * 0.1)




## Available Materials

The following materials are available for the rocket components:
aluminum, composite, fiberglass, carbon_fiber, balsa_wood, plywood, ABS_plastic

## Available Motors
Name,Manufacturer,Radius (mm),Length (mm),Dry Mass (kg),Max Thrust (N),Avg Thrust (N),Burn Time (s),Total Impulse (Ns),Isp (s), Cost ($)
Pro75M1670,CTI,75,757,1.815,2200,1533.9,3.9,6023.6,198, 520
AeroTechK700W,AT,54,568,0.732,1029.3,658.7,3.5,2249,177.5, 180
CesaroniM1670,CTI,75,757,3.101,2200,1533.9,3.6,6023.6,198, 550
AeroTechH128W,AT,29,194,0.108,190.5,141.2,1.29,176.5,191.3, 65
CesaroniO3700,CTI,161,957,14.194,4030.3,2836.9,8.2,29930.2,177.8, 1250
CesaroniO5800,CTI,150,754,12.418,6395.5,5040.2,5.2,30382.7,222, 1100
CesaroniK160,CTI,54,404,0.7,272.2,190.2,9.7,1521.7,182.9, 130

## Design Task

Based on the requirements and available components, design a rocket that will reach the target apogee. Your design should include:

1. Motor selection (choose from the available motors list)
2. Body dimensions and material
3. Nose cone dimensions and material
4. Fin design and material
5. Parachute specifications
6. Launch rail configuration


### Notes
DRC are run on the design so you need to make sure the design is feasible.
Here are some of the checks:
Notes for the tail the top and bottom radius cannot be the same (causes error)
The material must be specified exactly as listed above
The body radius must be greater than the motor radius
Nose cone but by exactly one of the listed 
Don't include any additional python code (other than config. Putting calculation in there is ok like 32/4 but not whole functions

## Response Format

Please provide your design as a Python dictionary that can be directly used in our simulation software. Use the following format:

```python
config = {
    "motor_choice": "MOTOR_NAME",  # Choose from available motors
    "rocket_body": {
        "radius": RADIUS_IN_METERS,  # Body radius in meters (must be greater than motor Radius)
        "length": LENGTH_IN_METERS,  # Body length in meters
        "material": "MATERIAL",  # Choose from available materials
        "thickness": THICKNESS_IN_METERS,  # Wall thickness in meters
    },
    "aerodynamics": {
        "nose_cone": {
            "kind": "SHAPE",  # "conical", "ogive", "elliptical", "tangent", "von karman", "parabolic", "powerseries" or "lvhaack".
            "length": LENGTH_IN_METERS,
            "material": "MATERIAL",
        },
        "fins": {
            "number": NUMBER_OF_FINS,
            "root_chord": LENGTH_IN_METERS,
            "tip_chord": LENGTH_IN_METERS,
            "span": LENGTH_IN_METERS,
            "cant_angle": ANGLE_IN_DEGREES,
            "material": "MATERIAL",
            "thickness": THICKNESS_IN_METERS,
        },
        "tail": {
            "length": LENGTH_IN_METERS,
            "top_radius": RADIUS_IN_METERS,
            "bottom_radius": RADIUS_IN_METERS,
            "material": "MATERIAL",
        },
    },
    "parachutes": {
        "main": {
            "name": "Main",
            "cd_s": AREA,
            "trigger": "apogee",
            "sampling_rate": 105,
            "lag": 1.5,
            "noise": (0, 8.3, 0.5),
        },
        "drogue": {
            "name": "Drogue",
            "cd_s": AREA,
            "trigger": "apogee",
            "sampling_rate": 105,
            "lag": 1.5,
            "noise": (0, 8.3, 0.5),
        },
    },
    "launch": {
        "rail_length": LENGTH_IN_METERS, 
        "inclination": ANGLE_IN_DEGREES, #90 is vertical launch
        "heading": ANGLE_IN_DEGREES, # Heading in degrees 0 is up
    },
    "payload": { #point mass as position specified 
        "mass": MASS_IN_KG,
        "position": POSITION_IN_METERS,  # relative to rocket center
    },
}
```


Here's an example valid design. This is not at all indicative of what you should do just an example:

```python
config = {
    "motor_choice": "CesaroniO5800",
    "rocket_body": {
        "radius": 0.1,  # Body radius in meters
        "length": 1.2,    # Body length in meters
        "material": "fiberglass",
        "thickness": 0.01,  # Wall thickness in meters
    },
    "aerodynamics": {
        "nose_cone": {
            "kind": "ogive",
            "length": 0.3,  # Nose cone length in meters
            "material": "composite",
        },
        "fins": {
            "number": 4,
            "root_chord": 0.15,  # Fin root chord in meters
            "tip_chord": 0.075,  # Fin tip chord in meters
            "span": 0.3,         # Fin span in meters
            "cant_angle": 0.5,   # Cant angle in degrees
            "material": "carbon_fiber",
            "thickness": 0.005   # Fin thickness in meters
        },
        "tail": {
            "length": 1.2,  # Tail length in meters
            "top_radius": 0.04,  # Top radius in meters
            "bottom_radius": 0.05,  # Bottom radius in meters
            "material": "carbon_fiber",
        },
    },
    "parachutes": {
        "main": {
            "name": "Main",
            "cd_s": 0.25,  # Main parachute CD_s
            "trigger": "apogee",
            "sampling_rate": 105,
            "lag": 1.5,
            "noise": (0, 8.3, 0.5),
        },
        "drogue": {
            "name": "Drogue",
            "cd_s": 0.2,  # Drogue parachute CD_s
            "trigger": "apogee",
            "sampling_rate": 105,
            "lag": 1.5,
            "noise": (0, 8.3, 0.5),
        },
    },
    "launch": {
        "rail_length": 1.2,  # Length of the launch rail in meters
        "inclination": 90,   # Rail inclination in degrees (vertical)
        "heading": 0,        # Launch heading in degrees (straight up)
    },
    "payload": {
        "mass": 0.5,  # Payload mass in kg
        "position": 0.6  # Payload position relative to rocket center in meters
    }
}
```
Before answering you should provide your full reasoning for the design choices you made thinking like a rocket scientist. Run sample calulations, make approximations etc to find best design
\end{lstlisting}






\end{document}